\documentclass[a4paper]{jpconf}
\usepackage{graphicx}

\begin{document}

\title{Three-nucleon forces and nuclei at the extremes}

\author{A Schwenk}

\address{Institut f\"ur Kernphysik, Technische Universit\"at Darmstadt, 
64289 Darmstadt, Germany and\\
ExtreMe Matter Institute EMMI, GSI Helmholtzzentrum f\"ur
Schwerionenforschung GmbH, 64291 Darmstadt, Germany}

\ead{schwenk@physik.tu-darmstadt.de}

\begin{abstract}
Neutron-rich nuclei become increasingly sensitive to three-nucleon
forces. These components of nuclear forces are at the forefront of
theoretical developments based on effective field theories of quantum
chromodynamics. We discuss our understanding of three-nucleon forces
and their impact on exotic nuclei, and show how new measurements test
and constrain them. Three-nucleon forces therefore provide an exciting
link between theoretical and experimental nuclear physics frontiers.
\end{abstract}

\section{Chiral effective field theory (EFT) and three-nucleon (3N) forces}

\begin{figure}[t]
\begin{center}
\includegraphics[trim=23mm 18mm 149mm 29mm,clip,width=0.44\textwidth]%
{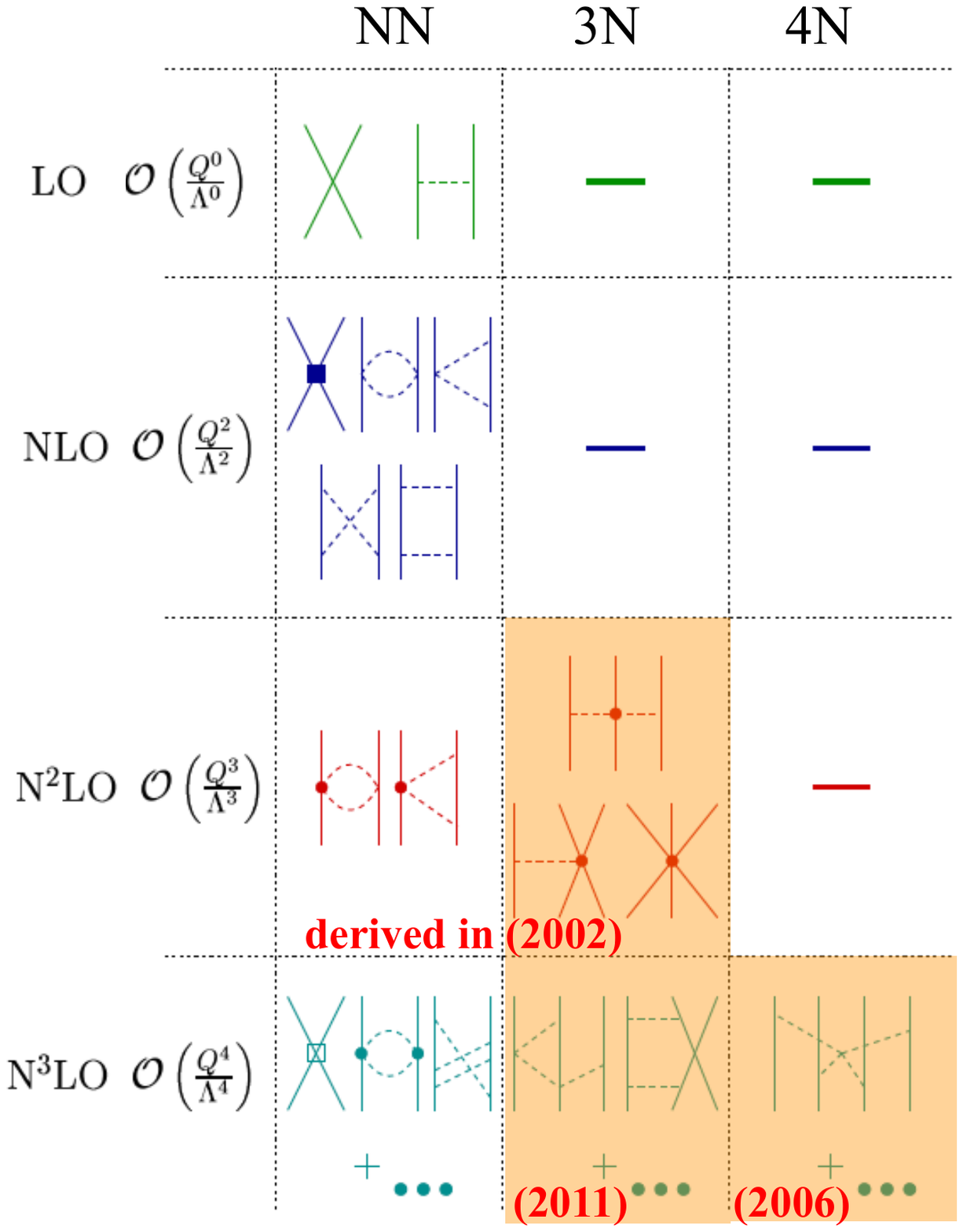}
\caption{Chiral EFT for nuclear forces, where the different contributions
at successive orders are shown diagrammatically~\cite{RMPEFT}. Many-body
forces are highlighted in orange including the year they were derived.
For neutrons, 3N and 4N forces are predicted parameter-free to N$^3$LO.
\label{chiralEFT}}
\end{center}
\end{figure}

Chiral EFT is based on the symmetries of quantum chromodynamics and is
applicable at momentum scales of order of the pion mass $Q \sim
m_\pi$, where pions are included as explicit degrees of freedom and
build up the long-range parts of strong interactions. In chiral EFT,
nucleons interact via pion exchanges and shorter-range contact
interactions~\cite{RMPEFT}. The resulting nuclear forces are organized
in a systematic expansion in powers of $Q/\Lambda_{\rm b}$, where
$\Lambda_{\rm b} \sim 500 \, {\rm MeV}$ denotes the breakdown
scale. As shown in Fig.~\ref{chiralEFT}, at a given order this
includes contributions from one- or multi-pion exchanges and from
contact interactions, with short-range couplings that are fit to data
and thus capture all short-range effects relevant at low energies. In
addition, nuclear forces depend on a resolution scale, where the
evolution is governed by the renormalization group (RG). The RG
decouples low and high momenta and leads to universal low-momentum
interactions with greatly enhanced convergence in few- and many-body
systems~\cite{PPNP}.

Chiral EFT opens up a systematic path to investigate many-body forces
and their impact on neutron-rich nuclei and neutron-rich
matter~\cite{RMP3N}. This results from the consistency of NN and 3N
interactions, which predicts the two-pion-exchange $c_1, c_3, c_4$
parts of 3N forces at N$^2$LO, leaving only two low-energy couplings
$c_D, c_E$ that encode pion interactions with short-range NN pairs and
short-range three-body physics. Moreover, all 3N and 4N forces at the
next order, N$^3$LO, are predicted~\cite{RMPEFT}. For systems of only
neutrons, the $c_D, c_E$ parts do not contribute because of the Pauli
principle and the coupling of pions to spin~\cite{nm}. Therefore,
chiral EFT predicts all three-neutron and four-neutron forces to
N$^3$LO. At the same time, 3N forces are a frontier in the physics of
nuclei and nucleonic matter in stars. This leads to a forefront
connection of 3N forces with the exploration of exotic nuclei at rare
isotope beam facilities worldwide.

\begin{figure}[t]
\begin{center}
\includegraphics[trim=8mm 16mm 4mm 88mm,clip,width=0.95\textwidth]{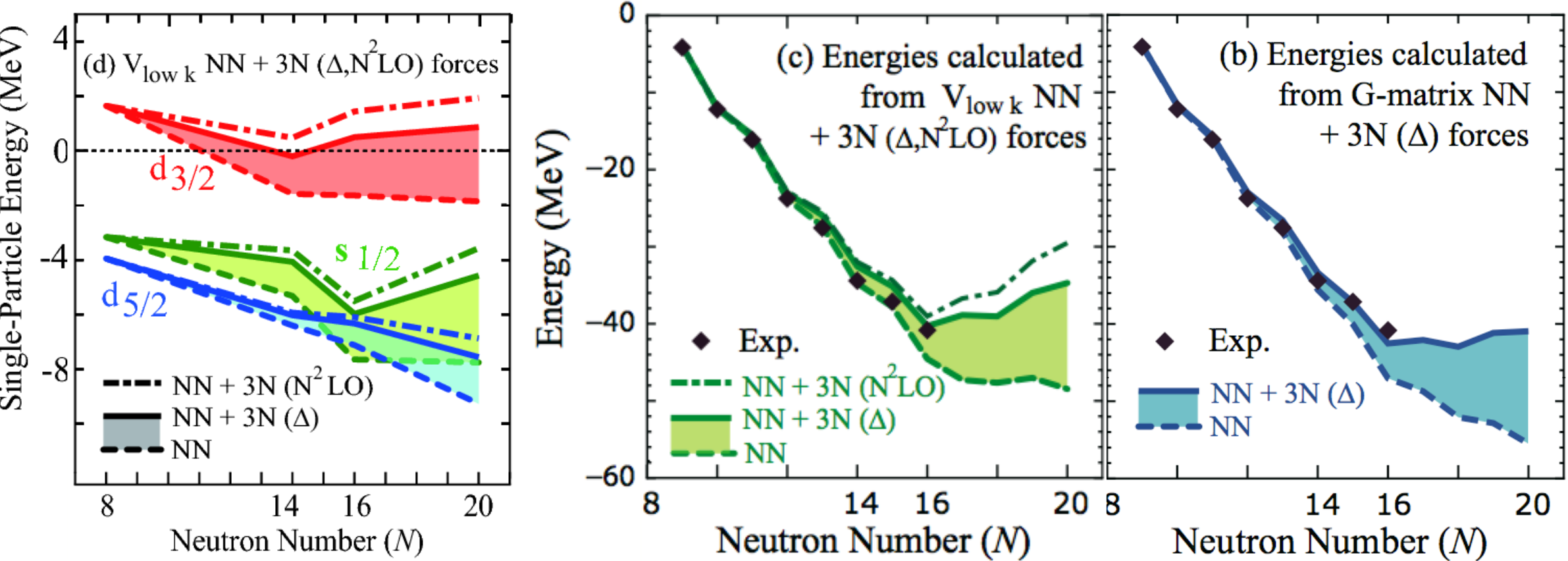}
\caption{Left panel: Single-particle energies in the oxygen isotopes
as a function of neutron number. Results are shown based on NN forces
only (RG-evolved to low-momentum interactions $V_{{\rm low}\,k}$) and
with N$^2$LO 3N forces (NN+3N). The changes due to the single-$\Delta$
contribution to 3N forces are highlighted by the shaded areas in all
panels. Right panels: Ground-state energies of the neutron-rich oxygen
isotopes relative to $^{16}$O, compared to the experimental energies
of the bound isotopes $^{17-24}$O. The middle panel shows the results
corresponding to the left panel. The right most panel is for a $G$
matrix. For details see Ref.~\cite{Oxygen}.\label{O}}
\end{center}
\end{figure}

\begin{figure}[t]
\begin{center}
\includegraphics[scale=0.34,clip=]{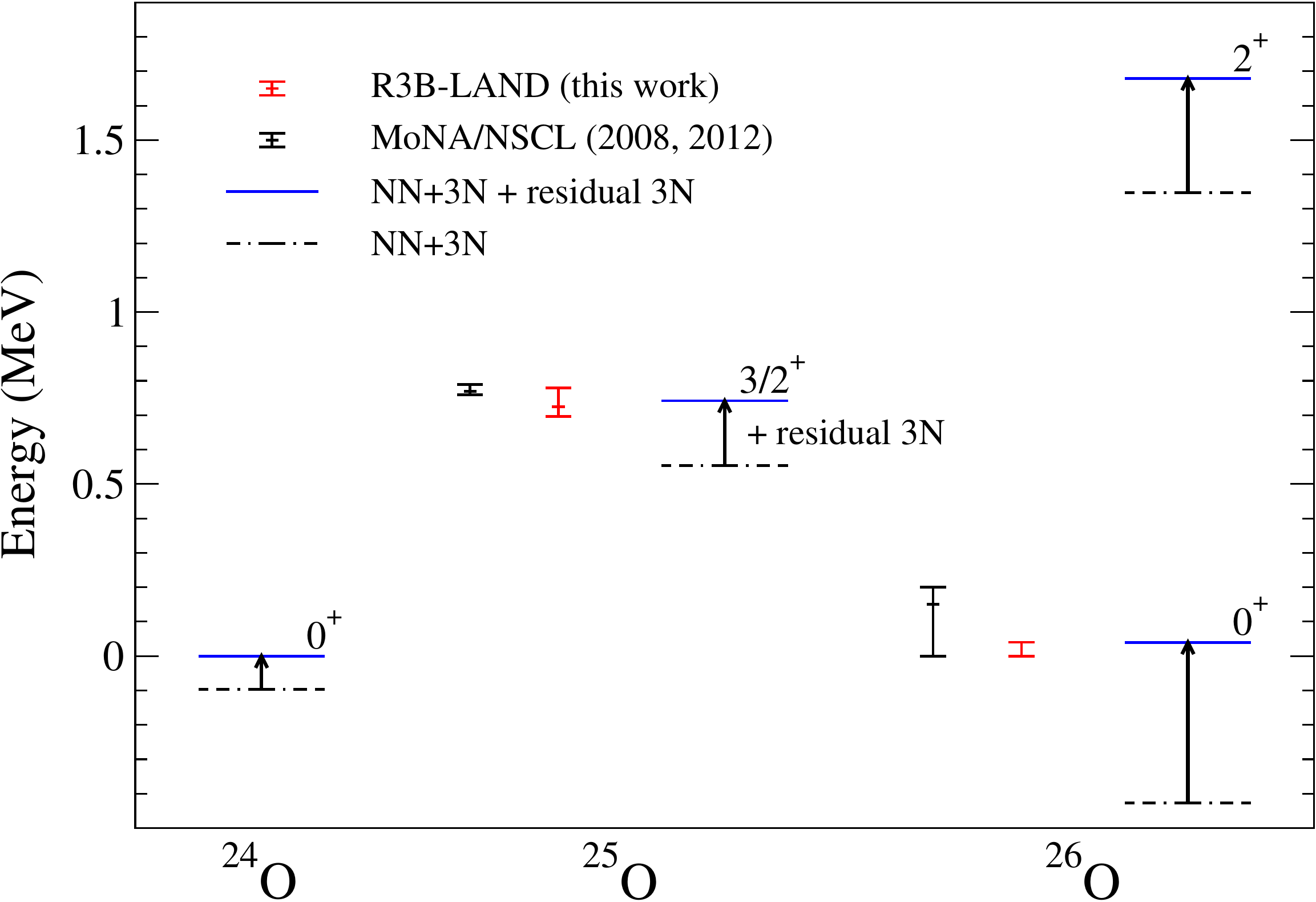}
\end{center}
\caption{Comparison of the experimental $^{25,26}$O energies
from MoNA~\cite{MoNA} and R3B-LAND~\cite{Caesar} with
theoretical shell-model calculations based on NN+3N forces and
including residual 3N forces. The impact of residual 3N forces is
highlighted by the arrows. For details see Ref.~\cite{Caesar}.
\label{residual}}
\end{figure}

\section{Three-nucleon forces and exotic nuclei}

Three-nucleon forces play a key role for understanding and predicting
exotic nuclei and for the formation and evolution of shell structure.
As shown in Fig.~\ref{O}, chiral 3N forces (fit only to $^3$H and
$^4$He) lead to repulsive interactions between valence neutrons that
change the location of the neutron dripline from $^{28}$O (with NN
forces only) to the experimentally observed $^{24}$O~\cite{Oxygen,Oxygen2}.
The position of the neutron dripline is driven by the location of the
$d_{3/2}$ orbital, which remains unbound with 3N forces. This presents
the first explanation of the oxygen anomaly based on nuclear
forces. The 3N-force mechanism is dominated by the single-$\Delta$
contribution (see the shaded areas in Fig.~\ref{O}) and was recently
confirmed in large-space calculations~\cite{CCO,Heiko,Carlo}.

\begin{figure}[t]
\begin{center}
\includegraphics[scale=0.68,clip=]{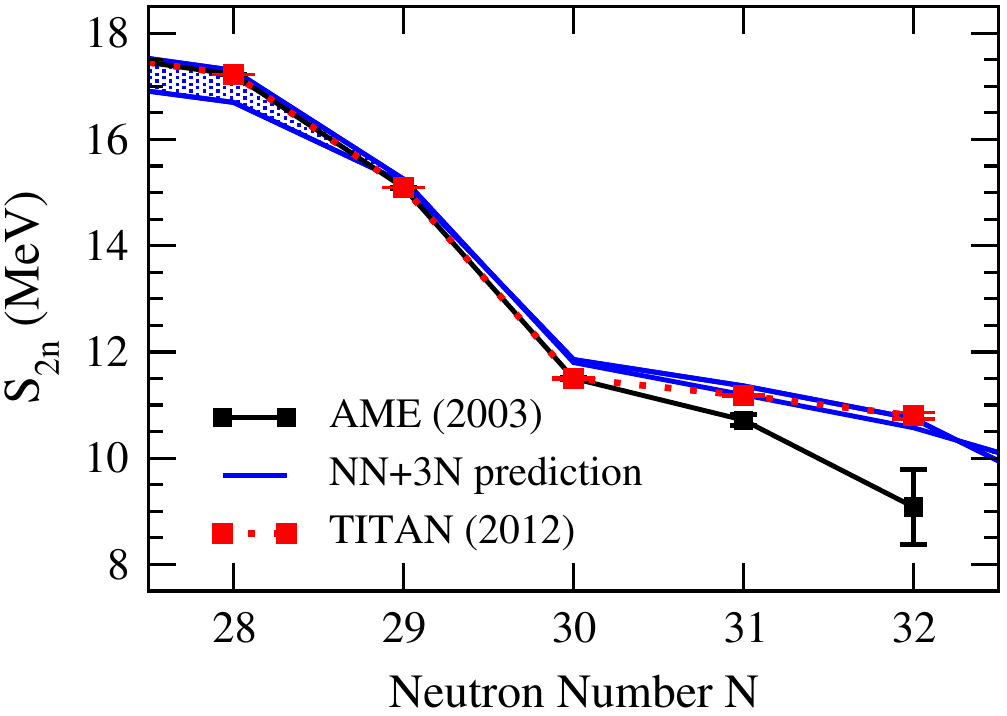}
\hspace*{2mm}
\includegraphics[scale=0.6,clip=]{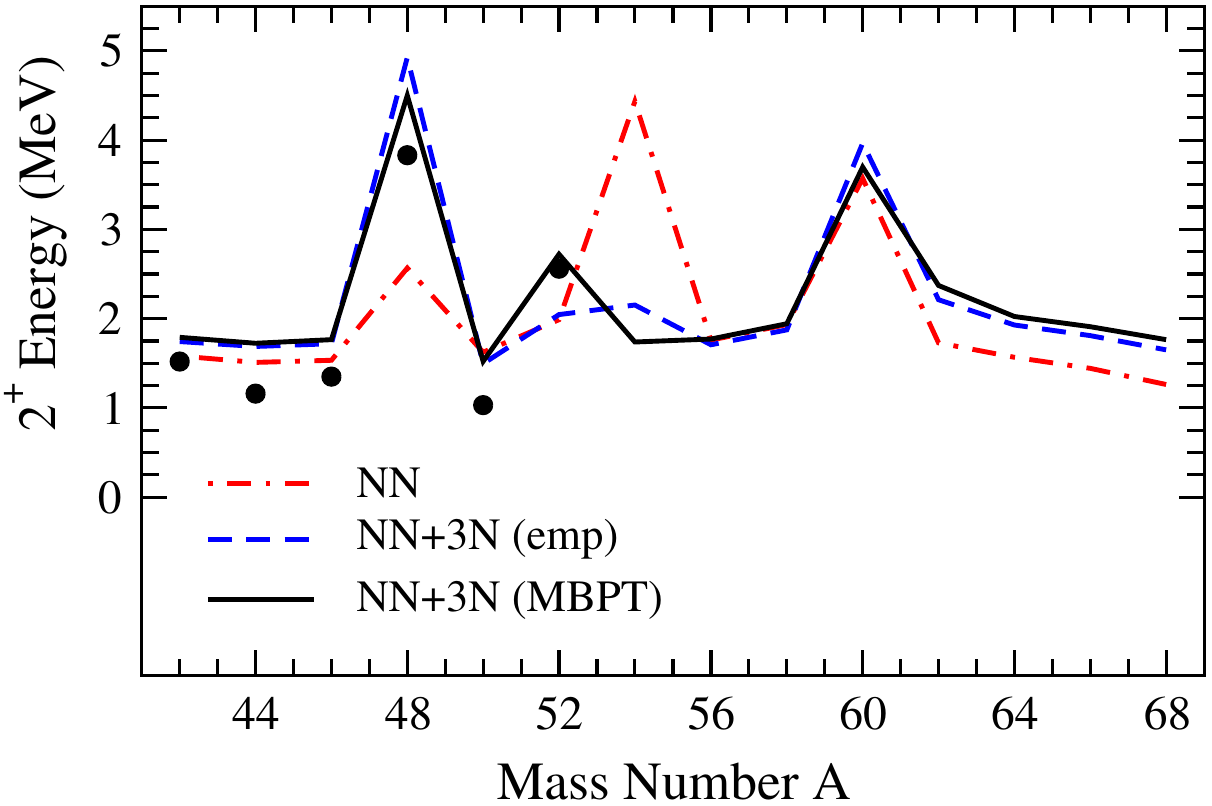}
\caption{Left panel: Two-neutron separation energy $S_{2n}$ of 
the neutron-rich calcium isotopes, with experimental energies from the AME
2003 atomic mass evaluation. We also show the new precision mass
measurements for $^{51,52}$Ca from TITAN, which disagree significantly
with the indirectly measured masses of AME 2003. Our predictions based
on NN+3N forces are in excellent agreement with these masses and with
the flat $S_{2n}$ behavior from $^{50}$Ca to $^{52}$Ca. For details
see Ref.~\cite{TITAN}. Right panel: $2^+$ energy in the even calcium
isotopes with and without 3N forces compared with experiment (dots
from ENSDF). The excitation energies are calculated to $^{68}$Ca in an
extended $pfg_{9/2}$ valence space, using both empirical (emp) and
calculated (MBPT) single-particle energies. For details see
Ref.~\cite{pairing}.\label{Ca}}
\end{center}
\end{figure}

\begin{figure}[t]
\begin{center}
\includegraphics[scale=0.54,clip=]{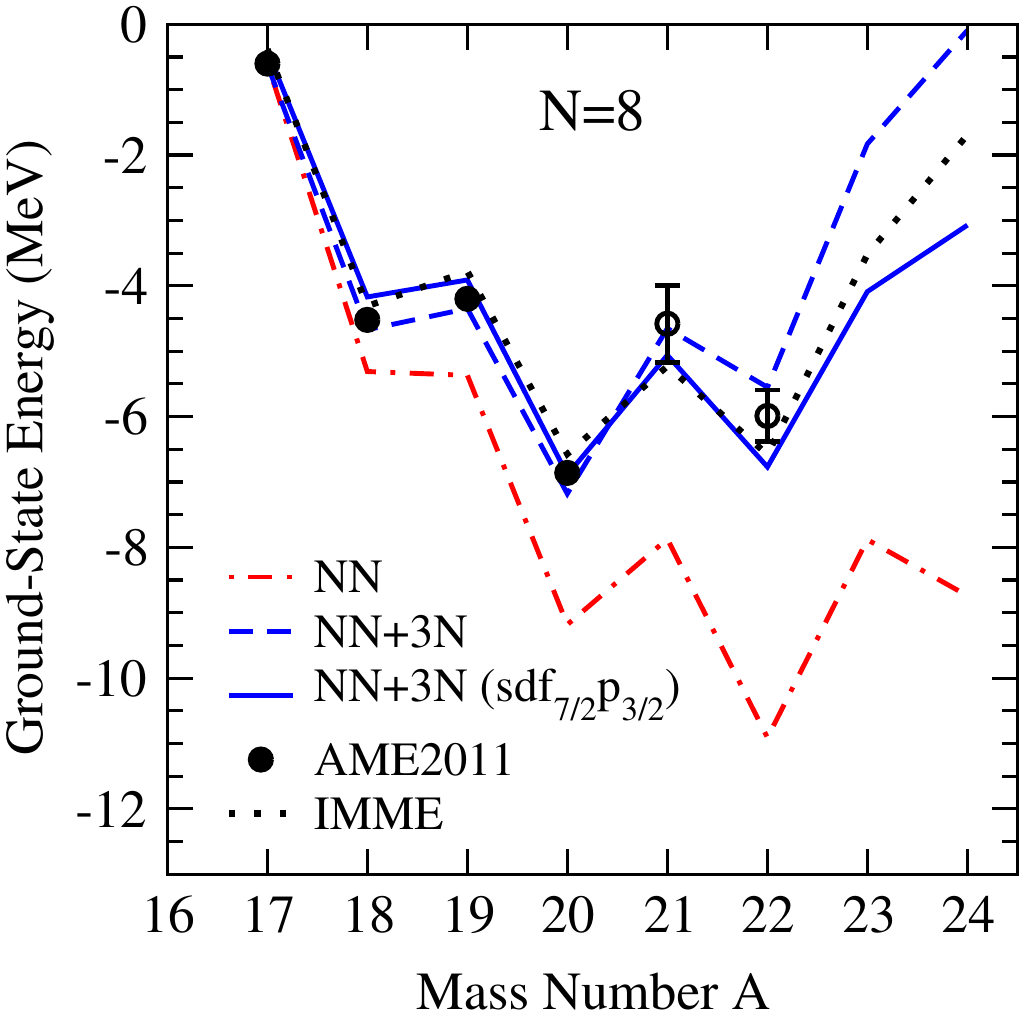}
\hspace*{2mm}
\includegraphics[scale=0.36,clip=]{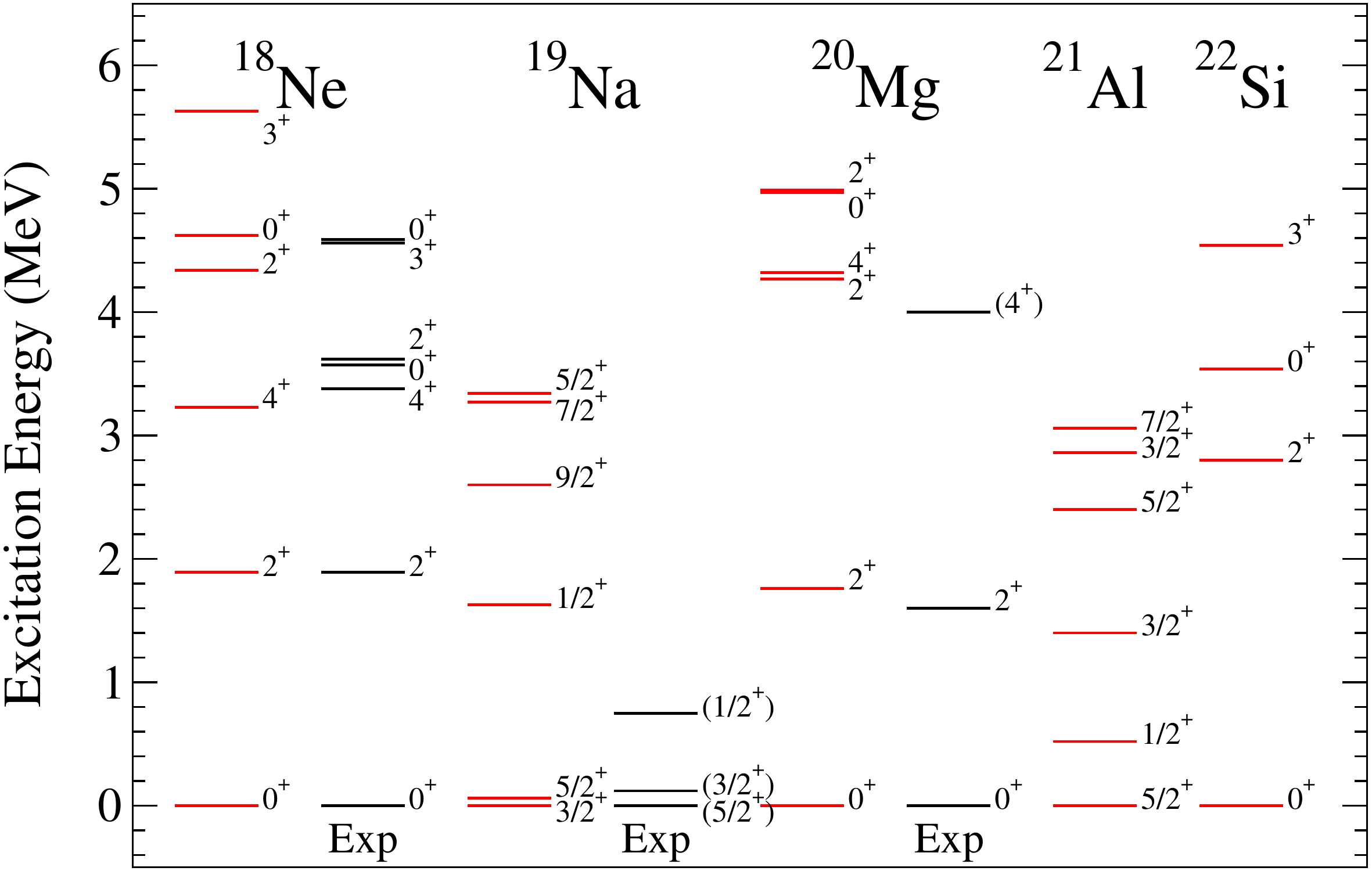}
\caption{Left panel: Ground-state energies of $N=8$ isotones relative
to $^{16}$O. Experimental energies (AME2011 with extrapolations as
open circles) and energies from the isobaric multiplet mass equation (IMME)
are shown. NN-only results in the $sd$ shell are compared to
calculations based on NN+3N forces in both $sd$ and $sdf_{7/2}p_{3/2}$
valence spaces with calculated (MBPT) single-particle energies. Right
panel: Excitation energies of $N=8$ isotones calculated with NN+3N
forces in the $sdf_{7/2}p_{3/2}$ valence space, compared with
experimental data where available. For details see
Ref.~\cite{protonrich}.\label{N8}}
\end{center}
\end{figure}

The interactions between valence neutrons are dominated by 3N forces
between two valence neutrons and one nucleon in the core of $^{16}$O.
This is expected for normal Fermi systems~\cite{Fermi}, where the
contributions from residual three-valence-nucleon interactions are
small compared to the normal-ordered two-body part. In the shell
model, the impact of residual 3N forces increases with the number of
valence nucleons, so they are amplified in the most neutron-rich
$^{25,26}$O isotopes studied at MoNA~\cite{MoNA} and
R3B-LAND~\cite{Caesar}, as demonstrated in Fig.~\ref{residual}.

While the magic numbers $N=2,8,20$ are generally well understood,
$N=28$ is the first standard magic number that is not reproduced in
microscopic theories with NN forces only. In studies for calcium
isotopes~\cite{Calcium,CCCa}, it was shown that 3N forces are key to
explain the $N=28$ magic number, leading to a high $2^+$ excitation
energy. Moreover, chiral 3N forces improve the agreement with
experimental masses, and as shown in Fig.~\ref{Ca} predicted a flat
behavior of the two-neutron separation energy from $^{50}$Ca to
$^{52}$Ca, in excellent agreement with new precision TITAN
Penning-trap mass measurements~\cite{TITAN}. The $2^+$ excitation
energy in the even calcium isotopes with and without 3N forces, based
on the same calculations as for the masses, is shown in
Fig.~\ref{Ca}~\cite{pairing}. This predicts a $2^+$ energy in
$^{54}$Ca of $1.7 - 2.2 \, {\rm MeV}$ (with 3N forces to first order
only, the $2^+$ energy is higher~\cite{Calcium}). The first
measurement of the $2^+$ energy in $^{54}$Ca was achieved at RIKEN and
presented by D.~Steppenbeck at this symposium.  Finally, we have
presented first results with 3N forces for the ground and excited
states of proton-rich nuclei along the $N=8$ (see Fig.~\ref{N8}) and
$N=20$ isotones to the proton dripline~\cite{protonrich}.

\begin{figure}[t]
\begin{center}
\includegraphics[scale=0.36,clip=]{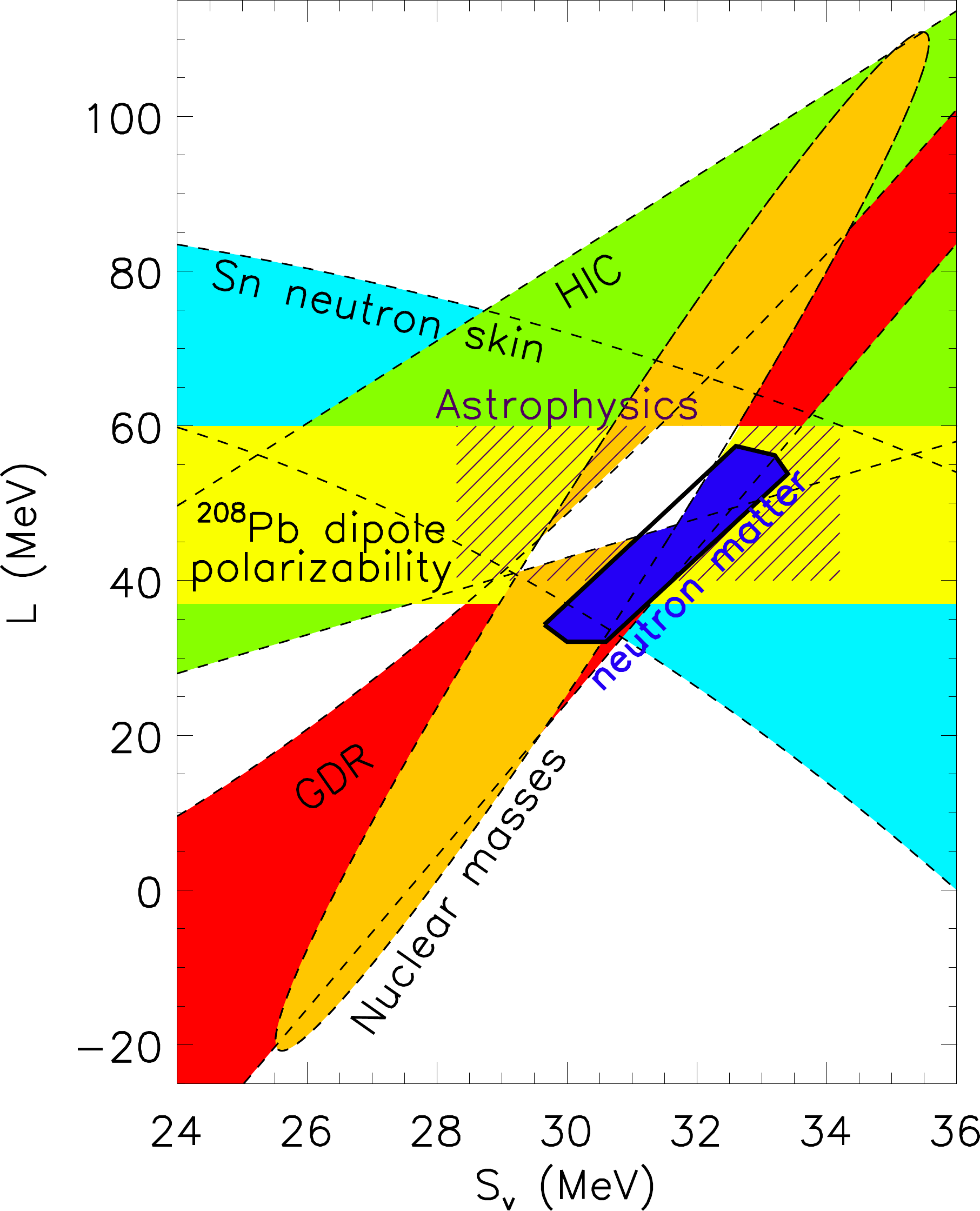}
\caption{Constraints for the symmetry energy $S_v$ and its density
derivative $L$. The blue region shows our neutron-matter constraints,
in comparison to bands from different empirical extractions (the white
area gives the overlap region)~\cite{LL}. For details see 
Ref.~\cite{nstarlong}.\label{sym}}
\end{center}
\end{figure}

\begin{figure}[t]
\begin{center}
\includegraphics[scale=0.34,clip=]{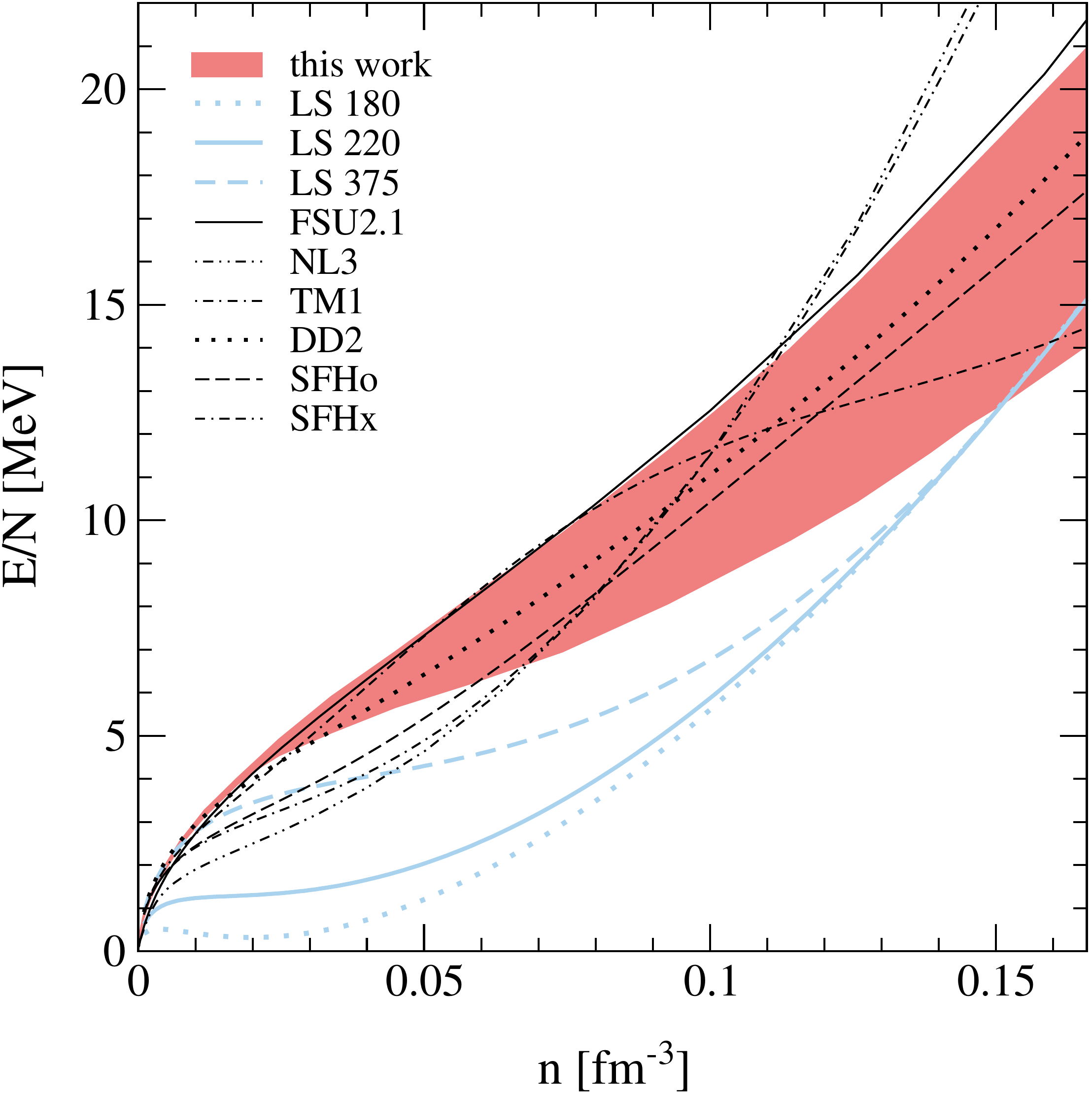}
\hspace*{2mm}
\includegraphics[scale=0.34,clip=]{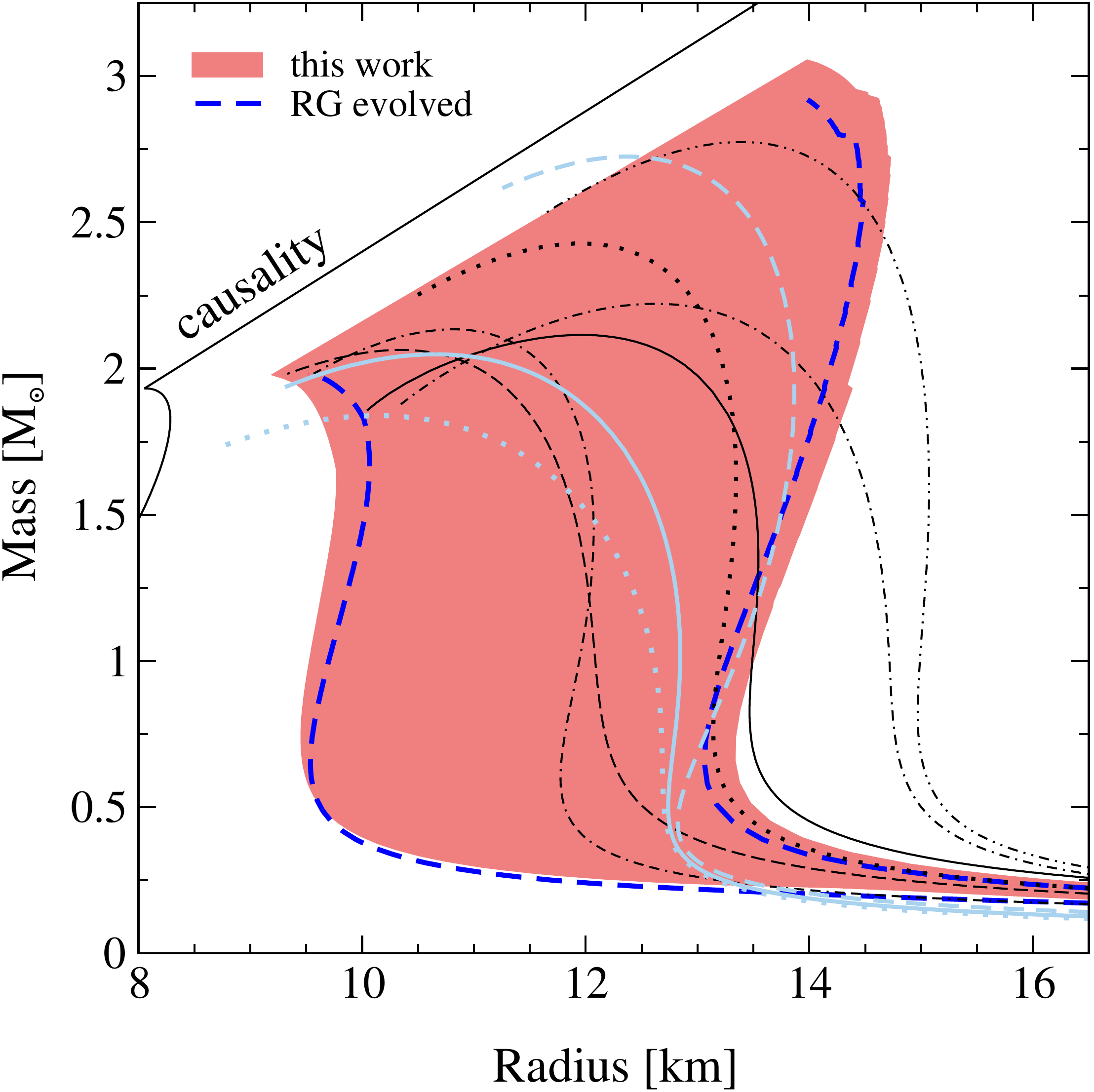}
\caption{Comparison of the neutron-matter energy at N$^3$LO (red band)
with equations of state for core-collapse supernova simulations provided
by Lattimer-Swesty (LS with different incompressibilities 180, 220,
and $375 \, {\rm MeV}$), G.~Shen (FSU2.1, NL3), Hempel (TM1, SFHo, SFHx),
and Typel (DD2). Right panel: Constraints on the mass-radius diagram
of neutron stars based on the neutron-matter results of the left panel
and following Ref.~\cite{nstarlong} for the extension to neutron-star
matter and to high densities (red band), in comparison to the
constraints based on RG-evolved interactions (thick dashed blue
lines)~\cite{nstarlong}. We also show the mass-radius relations
obtained from the equations of state for core-collapse supernova
simulations shown in the left panel (the same legend applies). For
details and references see Ref.~\cite{N3LOlong}.\label{neutmatt}}
\end{center}
\end{figure}

\section{Neutron matter and neutron stars}

The physics of neutron-rich matter ranges from universal properties at
low densities and in ultracold atoms to the densest matter we know to
exist in neutron stars. The same chiral 3N forces of the previous
section are repulsive in neutron matter and dominate the theoretical
uncertainties of the neutron-matter energy~\cite{nm}. The predicted
energy range provides tight constraints for the symmetry energy (see
Fig.~\ref{sym}) and predicts the neutron skin thickness of $^{208}$Pb
to $0.17 \pm 0.03 \, {\rm fm}$~\cite{nstar}, in excellent agreement
with a recent determination from the complete electric dipole
response~\cite{Tamii}. In addition, our calculations based on chiral
EFT interactions constrain the properties of neutron-rich matter below
nuclear densities to a much higher degree than is reflected in current
neutron star modeling~\cite{nstar}. These constraints have been recently
explored for the gravitational wave signal in neutron star
mergers~\cite{Bauswein}.

To improve our understanding of neutron matter further, we have
performed the first complete N$^3$LO calculation of neutron matter
including NN, 3N and 4N forces~\cite{N3LO,N3LOlong}. The resulting
energy is shown in the left panel of Fig.~\ref{neutmatt}. This leads
to bands consistent with the RG-evolved results~\cite{nm}. Many of the
equations of state for core-collapse supernova simulations are
inconsistent with the N$^3$LO neutron-matter band.  Combined with the
heaviest $2 M_\odot$ neutron star~\cite{Demorest}, our results shown
in Fig.~\ref{neutmatt} constrain the radius of a typical $1.4 M_\odot$
star to $R=9.7-13.9 \, {\rm km}$~\cite{nstarlong,N3LOlong} (the same
relative uncertainty as the neutron skin). The predicted radius range
is due, in about equal amounts, to the uncertainty in 3N forces and to
the extrapolation to high densities. The radius range is also
consistent with astrophysical results obtained from modeling X-ray
burst sources~\cite{Steiner}.  The physics of 3N forces therefore
connects the heaviest neutron-rich nuclei with the heaviest neutron
stars.

\section*{Acknowledgments}

All the very best for your 60th birthday, Taka, lots of good health,
happiness and continued success! It has been a great pleasure working
together on 3N forces and exotic nuclei and I look forward to exciting
future work. I would also like to thank J.~Dilling, A.~T.~Gallant,
K.~Hebeler, J.~D.~Holt, T.~Kr\"uger, J.~M.~Lattimer, J.~Men{\'e}ndez,
C.~J.~Pethick, J.~Simonis, T.~Suzuki, and I.~Tews, who contributed to
the results presented in this talk. This work was supported by the
BMBF under Contract No.~06DA70471, the DFG through Grant SFB 634, the
ERC Grant No.~307986 STRONGINT, and the Helmholtz Alliance HA216/EMMI.

\end{document}